\documentclass[pra,twocolumn,showpacs,floatfix]{revtex4-1}
\usepackage{amsmath}
\usepackage{amssymb}
\usepackage{graphicx}
\usepackage[english]{babel}
\usepackage{latexsym}
\usepackage{graphics}
\usepackage{subfigure}

\def\be{\begin{equation}}
\def\ee{\end{equation}}
\def\bea{\begin{eqnarray}}
\def\eea{\end{eqnarray}}
\def\bse{\begin{subequations}}
\def\ese{\end{subequations}}

\renewcommand{\v}[1]{{\bf #1}}

\def\be{\begin{eqnarray}}
\def\ee{\end{eqnarray}}

\begin{document}

\title{Spin-orbit coupling and perpendicular Zeeman field for fermionic cold
atoms: Observation of the intrinsic anomalous Hall effect}
\author{Chuanwei Zhang$^{1}$}

\begin{abstract}
We propose a scheme for generating Rashba spin-orbit coupling and
perpendicular Zeeman field \textit{simultaneously} for cold fermionic atoms
in a harmonic trap through the coupling between atoms and laser fields. The
realization of Rashba spin-orbit coupling and perpendicular Zeeman field
provides opportunities for exploring many topological phenomena using cold
fermionic atoms. We focus on the intrinsic anomalous Hall effect and show
that it may be observed through the response of atomic density to a rotation
of the harmonic trap.
\end{abstract}
\affiliation{$^{1}$Department of Physics and Astronomy, Washington State University,
Pullman, Washington, 99164 USA}
\maketitle

Two important ingredients for manipulating electron spin dynamics and
designing spin devices in spintronics \cite{Sarma} are spin-orbit coupling
and Zeeman field. For instance, Rashba spin-orbit coupling (RSOC) (assume in
the $xy$ plane), together with a \textit{perpendicular} Zeeman field (PZF)
(along $\mathbf{e}_{z}$), yield a transverse (along $\mathbf{e}_{y}$)
topological Hall current with an applied electric field (along $\mathbf{e}%
_{x}$). Such an intrinsic current was proposed to be one feasible
explanation of the experimentally observed anomalous Hall effect (AHE) and
spin Hall effect (SHE) in ferromagnetic semiconductors \cite{Niu1,Niu3}.
However, the scattering of electrons from impurities and defects in the
solid, leading to extrinsic AHE and SHE, makes the experimental observation
of intrinsic AHE and SHE very difficult \cite{Nagaosa}.

Ultra-cold atomic gases experience an environment essentially free from
impurities and defects, and therefore provide an ideal platform to emulate
many condensed matter models or even observe new phenomena. One important
recent effort along this line is in study of atomtronics that, in analogy to
electronics and spintronics, aims to realize devices and circuits using cold
atoms \cite{Holland}. One natural and important question in atomtronics is
how to generate effective spin-orbit coupling and Zeeman fields. Great
progress has been made recently on the generation of RSOC by considering the
coupling between cold atoms and laser fields \cite{Ruseckas,Zhu,Liu,Liu2},
which leads to a series of important applications \cite%
{Ruseckas2,Clark1,Clark2,Zhang}.

However, the direction of the generated Zeeman field in these previous
schemes \cite{Ruseckas,Liu2,Ruseckas2,Clark1,Clark2,Zhang} is in the
spin-orbit coupling plane. Such \textit{in-plane} Zeeman fields cannot open
a band gap between different energy branches in the energy spectrum. The
band gap, together with RSOC, is the physical origin of many topological
phenomena. For instance, the band gap is necessary for the observation of
the intrinsic AHE \cite{Niu1}. It is also the key ingredient of the recently
broadly discussed schemes on the creation of a chiral $p_{x}+ip_{y}$-wave
superfluid/superconductor from an \textit{s}-wave superfluid/superconductor
\cite{Zhangcw,Sau} for the observation of non-Abelian statistics and
topological quantum computation \cite{Nayak}. In contrast, a band gap can be
opened in the presence of RSOC and PZF, but a scheme for generating them
simultaneously for cold atoms is still absent.

In this paper, we propose a scheme to create RSOC and PZF \textit{%
simultaneously} for a fermionic atomic gas in a harmonic trap. The
realization of RSOC and PZF may open opportunities for the observation of
many topological phenomena in cold atoms because of the non-zero Berry phase
induced by RSOC and PZF. Here we focus on one of them: observation of the
intrinsic AHE. In solid state systems, the AHE has been observed in
transport experiments for electrons (\textit{i.e.}, measuring charge
currents or voltages). Such transport experiments are not suitable for cold
atoms in a harmonic trap. We find that the time-of-flight of cold atoms in
the presence of RSOC and PZF can only yield a small asymmetry of the atomic
density, therefore it may not be suitable for observing the intrinsic AHE.
Instead, we consider the response of atom density to an external rotation of
the trap, which corresponds to an effective magnetic field for atoms.
Because the intrinsic AHE is not an quantized effect, the Str\u{e}da formula
\cite{Streda} that was proposed for studying quantum Hall effects in cold
atoms \cite{Zhai} does not apply. We find that the atomic density response
to the rotation contains not only contributions from the anomalous Hall
conductivity, but also a new term from the orbit magnetic moments of atoms
that is absent in previous literature \cite{Xiao}. Both contributions
originate from the topological properties of RSOC and PZF.

\begin{figure}[t]
\includegraphics[width=0.9\linewidth]{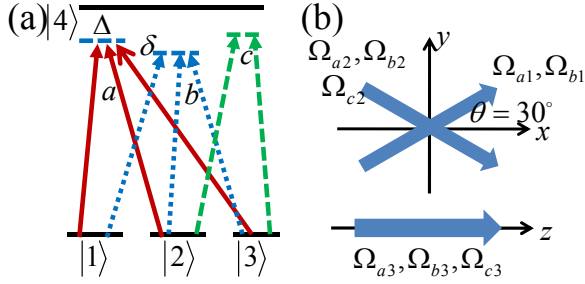} \vspace{-10pt}
\caption{(Color online) Schematic representation of the light-atom
interaction for the generation of the effective Hamiltonian (\protect\ref%
{Ham5}). $a$, $b$, $c$ are three sets of lasers. Soild: $\Omega _{ai}$;
Dotted: $\Omega _{bi}$; Dashed: $\Omega _{ci}$. $\Delta $ is the laser
detuning of laser set $a$, $\protect\delta $ is a small shift of the
detuning for laser sets $b$ and $c$ from $\Delta $. (b) The configuration of
the laser beams. All lasers are uniform plane waves.}
\label{laser}
\end{figure}

Consider ultra-cold Fermi atoms with a tripod electronic level scheme (Fig. %
\ref{laser}a). States $\left\vert 1\right\rangle $, $\left\vert
2\right\rangle $, $\left\vert 3\right\rangle $ are three hyperfine ground
states, and state $\left\vert 4\right\rangle $ is an excited state. The
fermi gas is confined in a quasi-two dimensional (\textit{xy}-plane)
harmonic trap. Along the \textit{z} direction, the atomic dynamics is
\textquotedblleft frozen\textquotedblright\ by a deep optical lattice,
leading to a multiple layered system. The ground states $\left\vert
1\right\rangle $, $\left\vert 2\right\rangle $, $\left\vert 3\right\rangle $
are coupled with state $\left\vert 4\right\rangle $ by three lasers with
corresponding Rabi frequencies $\Omega _{a1}$, $\Omega _{a2}$, and $\Omega
_{a3}$. In the interaction representation, the single-particle Hamiltonian is%
\begin{equation}
H=p^{2}/2m+V_{ext}+H_{I}-\mu ,  \label{Ham1}
\end{equation}%
where $H_{I}=\hbar \Delta \left\vert 4\right\rangle \left\langle
4\right\vert -\hbar (\Omega _{a1}\left\vert 4\right\rangle \left\langle
1\right\vert +\Omega _{a2}\left\vert 4\right\rangle \left\langle
2\right\vert +\Omega _{a3}\left\vert 4\right\rangle \left\langle
3\right\vert +H.c.)$ describes the laser-atom interaction, where $\Delta $
is the detuning to state $\left\vert 4\right\rangle $. $V_{ext}$ is the
external potential that includes the harmonic trap as well as potentials
created by other laser fields, $\mu $ is the chemical potential.

The Rabi frequencies $\Omega _{ai}$ can be parameterized as $\Omega
_{a1}=\Omega _{a}\sin \theta \cos \varphi e^{iS_{1}}$, $\Omega _{a2}=\Omega
_{a}\sin \theta \sin \varphi e^{iS_{2}}$, $\Omega _{a3}=\Omega _{a}\cos
\theta e^{iS_{3}}$, and $\Omega _{a}=\sqrt{\left\vert \Omega
_{a1}^{2}\right\vert +\left\vert \Omega _{a2}^{2}\right\vert +\left\vert
\Omega _{a3}^{2}\right\vert }$. The diagonalization of $H_{I}$ yields two
degenerate dark states: $\left\vert D_{1}\right\rangle =\sin \varphi
e^{iS_{31}}\left\vert 1\right\rangle -\cos \varphi e^{iS_{32}}\left\vert
2\right\rangle $, $\left\vert D_{2}\right\rangle =\cos \theta \cos \varphi
e^{iS_{31}}\left\vert 1\right\rangle +\cos \theta \sin \varphi
e^{iS_{32}}\left\vert 2\right\rangle -\sin \theta \left\vert 3\right\rangle $%
, with $S_{ij}=S_{i}-S_{j}$. We choose a laser configuration illustrated in
Fig. \ref{laser}b with the parameters $\varphi =\pi /4$, $\theta =\arctan
\sqrt{2}$, $S_{i}=\mathbf{q}_{i}\cdot \mathbf{R}$, $\mathbf{q}%
_{1}=q_{0}\left( \cos \frac{\pi }{6}\mathbf{e}_{x}+\sin \frac{\pi }{6}%
\mathbf{e}_{y}\right) $, $\mathbf{q}_{2}=q_{0}\left( \cos \frac{\pi }{6}%
\mathbf{e}_{x}-\sin \frac{\pi }{6}\mathbf{e}_{y}\right) $, $\mathbf{q}%
_{3}=q_{1}\mathbf{e}_{z}$, $\mathbf{R}=\left( x,y,z\right) $. The lasers $%
\Omega _{a1}$, $\Omega _{a2}$ are in the $xy$ plane, $\Omega _{a3}$ is along
the $z$ direction, and $\left\vert \Omega _{a1}\right\vert =\left\vert
\Omega _{a2}\right\vert =\left\vert \Omega _{a3}\right\vert $. Note that
these three lasers are uniform plane waves, which are different from the
optical lattices used in a previous scheme \cite{Zhang} to generate RSOC.
The effective low-energy Hamiltonian is obtained by projecting the
Hamiltonian (\ref{Ham1}) onto the subspace of the degenerate dark states
spanned by $\left\vert D_{i}\right\rangle $
\begin{equation}
H_{e}=\gamma k^{2}-\mu -\alpha \left( k_{x}\sigma _{z}+k_{y}\sigma
_{x}\right) +V,  \label{Ham2}
\end{equation}%
where $\gamma =\hbar ^{2}/2m$, $\alpha =\gamma q_{0}/\sqrt{3}$, $%
V_{ij}=\left\langle D_{i}\right\vert V_{ext}\left\vert D_{j}\right\rangle $
is the effective external potential. The external harmonic trap $%
V_{trap}=V\left( \mathbf{r}\right) \sum\nolimits_{i=1}^{3}\left\vert
i\right\rangle \left\langle i\right\vert $ is chosen to be spin-independent
to avoid heating of atoms, where $V\left( \mathbf{r}\right) =m\omega
_{t}^{2}r^{2}/2$, $\mathbf{r}=\left( x,y\right) $ is the coordinate in the $%
xy$ plane, $\omega _{t}$ is the trapping frequency.

Two additional laser sets $b$, $c$ with Rabi frequencies $\Omega _{bi}$, $%
\Omega _{ci}$ in Fig. \ref{laser} induce two Raman transitions between
different hyperfine ground states, yielding an effective coupling
interaction
\begin{equation}
H_{Z}=-\frac{\hbar }{\Delta }\left(
\begin{array}{ccc}
\left\vert \Omega _{b1}^{2}\right\vert & \Omega _{b1}^{\ast }\Omega _{b2} &
\Omega _{b1}^{\ast }\Omega _{b3} \\
\Omega _{b2}^{\ast }\Omega _{b1} & \digamma _{1} & \digamma _{3} \\
\Omega _{b3}^{\ast }\Omega _{b1} & \digamma _{3}^{\ast } & \digamma _{2}%
\end{array}%
\right)  \label{Ham3}
\end{equation}%
for atoms at the hyperfine ground states $\left\vert 1\right\rangle $, $%
\left\vert 2\right\rangle $, $\left\vert 3\right\rangle $, where $\digamma
_{1}=\left\vert \Omega _{b2}^{2}\right\vert +\left\vert \Omega
_{c2}^{2}\right\vert $, $\digamma _{2}=\left\vert \Omega
_{b3}^{2}\right\vert +\left\vert \Omega _{c3}^{2}\right\vert $, $\digamma
_{3}=\Omega _{b2}^{\ast }\Omega _{b3}+\Omega _{c2}^{\ast }\Omega _{c3}$. We
choose the detunings $\Delta _{b}$, $\Delta _{c}$ for the two sets ($b$ and $%
c$) of lasers as $\Delta _{b}=\Delta +\delta $, $\Delta _{c}=\Delta -\delta $%
, with $\delta \sim 2\pi \times 80$ MHz$\ll \Delta $. The small shifts of
the detunings do not change the effective Rabi coupling between different
hyperfine ground states, but remove the interference among different sets of
lasers. The optical potentials generated by the laser sets $b$ and $c$ are
taken as external potentials. With suitably chosen Rabi frequencies: $\Omega
_{b1}=i\sqrt{3}\Omega _{0}e^{iS_{1}}$, $\Omega _{b2}=-\Omega _{0}e^{iS_{2}}$%
, $\Omega _{b3}=\Omega _{0}e^{iS_{3}},\Omega _{c2}=\sqrt{2}e^{i\pi /3}\Omega
_{0}e^{iS_{2}}$, $\Omega _{c3}=\sqrt{2}\Omega _{0}e^{iS_{3}}$, the
Hamiltonian (\ref{Ham3}) reduces to $H_{Z}=ih_{0}\left( \left\vert
D_{2}\right\rangle \left\langle D_{1}\right\vert -\left\vert
D_{1}\right\rangle \left\langle D_{2}\right\vert \right) $ with $%
h_{0}=3\hbar \Omega _{0}^{2}/\Delta $. Here $\Omega _{0}$ is the magnitude
of the Rabi frequency of the laser $b2$. Within the dark state basis $%
\left\vert D_{i}\right\rangle $, $H_{Z}=h_{0}\sigma _{y}$. Under a new dark
state basis $\left\vert \chi _{1}\right\rangle =\left( \left\vert
D_{1}\right\rangle -i\left\vert D_{2}\right\rangle \right) /\sqrt{2}$, $%
\left\vert \chi _{2}\right\rangle =\left( -i\left\vert D_{1}\right\rangle
+\left\vert D_{2}\right\rangle \right) /\sqrt{2}$, which corresponds to a
unitary rotation of $\left\vert D_{i}\right\rangle $, the Hamiltonian (\ref%
{Ham2}) becomes
\begin{equation}
H_{e}=\gamma k^{2}-\mu +\alpha \left( k_{x}\sigma _{y}-k_{y}\sigma
_{x}\right) +h_{0}\sigma _{z}+V\left( \mathbf{r}\right) ,  \label{Ham5}
\end{equation}%
where the third term is the RSOC, the fourth term is the PZF. All eight
lasers used for the generation of RSOC and PZF are uniform plane waves,
therefore they do not lead to spatial periodic modulation of the atomic
density. In addition, these lasers propagate only along three different
directions (the same as other tripod schemes \cite%
{Ruseckas,Ruseckas2,Clark1,Zhang}), therefore our scheme should be feasible
in experiments.

\begin{figure}[t]
\includegraphics[width=0.9\linewidth]{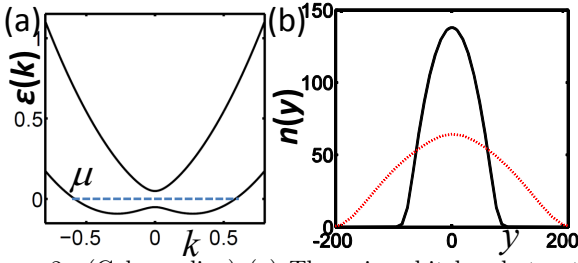} \vspace{-5pt} \vspace{-10pt}
\caption{(Color online) (a) The spin-orbit band structure. $h_{0}=0.05E_{R}$%
. $\protect\alpha ^{2}/\protect\gamma =1/3E_{R}$. We use the wavelength $%
\protect\lambda =767$ nm of the $^{40}K$ D$_{2}$ transition line, the
wavevector $k_{R}=2\protect\pi /\protect\lambda $, the recoil frequency $%
\protect\omega _{R}=\hbar k_{R}^{2}/2m=2\protect\pi \times 8.5KHz$, the
recoil energy $E_{R}=\hbar \protect\omega _{R}$, as the units of length,
wavevector, frequency, and energy, respectively. $E_{R}$ corresponds to a
temperature $T=2\protect\mu K$. (b) The time of flight image of atoms in the
presence of PSOC, PZF, and gravitational force. The unit of the density is $%
k_{R}^{2}=6.7\times 10^{9}$ cm$^{-2}$. $\protect\mu =0$, $T=20nK$, $\protect%
\omega _{t}=2\protect\pi \times 10Hz$. Solid: $t=0$; Dotted: $t=19$ ms.}
\label{TOF}
\end{figure}

Under the local density approximation with the local chemical potential $\mu
\left( \mathbf{r}\right) =\mu -m\omega _{t}^{2}r^{2}/2$, the Hamiltonian (%
\ref{Ham5}) has two eigenenergies $\varepsilon _{\mathbf{k}\pm }=\gamma
k^{2}-\mu \left( \mathbf{r}\right) \pm E_{0}$ with $E_{0}=\sqrt{%
h_{0}^{2}+\alpha ^{2}k^{2}}$. There is an energy gap $E_{g}=2h_{0}$ opening
between two spin orbit bands at $k=0$ (Fig. \ref{TOF}a). The intrinsic AHE
is nonzero only when the chemical potential $\mu \left( \mathbf{r}\right) $
lies inside the gap.

The dynamics of cold fermi atoms are described by the semiclassical
equations of motion \cite{Niu4}
\begin{equation}
\mathbf{\dot{r}}=\partial \varepsilon _{\mathbf{k}}/\partial \mathbf{k}-%
\mathbf{\dot{k}\times \Gamma }_{z}\text{, }\mathbf{\dot{k}}=\mathbf{F/\hbar ,%
}  \label{EOM1}
\end{equation}%
where $\Gamma \left( \mathbf{k}\right) =\alpha ^{2}h_{0}/2\left( \alpha
^{2}k^{2}+h_{0}^{2}\right) ^{3/2}\mathbf{e}_{z}$ is the Berry curvature in
the momentum space and the physical origin of many topological phenomena.
When there is a nonzero external force $\mathbf{F}$ along the $x$ direction,
an anomalous velocity $\mathbf{v}_{AH}=-\mathbf{\dot{k}\times \Gamma }$ of
atoms is induced along the $y$ direction. $\mathbf{v}_{AH}$ is the physical
origin of the intrinsic AHE in electronic systems. However, it is difficult
to perform a similar transport measurement commonly used for electronic
systems for cold atoms in a harmonic trap. In the schemes for the
observation of SHE in cold atoms \cite{Zhu,Liu,Liu2}, the time-of-flight
image has been proposed. Because of the anomalous velocity $\mathbf{v}_{AH}$%
, we expect the time-of-flight expansion of cold atoms with RSOC, PZF and
the gravitational force $\mathbf{F}=m\mathbf{g}$ (along the $x$ direction)
has a transverse shift along the $y$ direction.

We assume that the harmonic trap in the $xy$ plane is suddenly turned off at
$t=0$, and atoms start to expand, following the equations of motion (\ref%
{EOM1}). The column density along $y$ direction after the expansion depends
on the initial atom distribution in both momentum and real spaces
\begin{equation}
n\left( y_{f},T,t\right) =\int d^{2}\mathbf{r}_{i}\int \frac{d^{2}\mathbf{k}%
}{\left( 2\pi \right) ^{2}}\delta \left( y_{f}-y_{i}-\Delta y\left( t\right)
\right) f,  \label{denTOF}
\end{equation}%
where $T$ is the temperature, $t$ is the time of flight. $\mathbf{r}_{i}$ ($%
y_{i}$) is the initial position of atoms, and $f\left( \mathbf{k},\mathbf{r}%
_{i},T\right) =1/\left( \exp \left[ \left( \varepsilon _{\mathbf{k}}-\mu
\left( \mathbf{r}_{i}\right) \right) /k_{B}T\right] +1\right) $ is the
Fermi-Dirac distribution of atoms. $y_{f}$ is the position of atoms at time $%
t$. $\Delta y\left( t\right) =\int_{k_{x}}^{k_{x}+Ft/\hbar }dk_{x}^{\prime
}\left( \frac{1}{F}\frac{\partial \varepsilon _{\mathbf{k}}}{\partial k_{y}}%
+\Gamma _{z}\left( k_{x}^{\prime },k_{y}\right) \right) $ is the distance of
flight. $\delta \left( y\right) $ is the delta function.

In Fig. \ref{TOF}b, we plot the column density $n\left( y_{f},T,t\right) $
at two different times. At $t=0$, the initial column density is symmetric
along the $y$ axis. After a time $t$, $n\left( y_{f},T,t\right) $ becomes
asymmetric because of the anomalous velocity of atoms. However, the
expansion dynamics of atoms is dominated by the first term in $\Delta
y\left( t\right) $, and the differences of the atomic densities at $\pm
y_{f} $ correspond to only a small percentage (below 3\%) of the total
density. Therefore it may be hard to observe them in experiments. The
difference between electrons and cold atoms comes from the fact that $%
\partial \varepsilon _{\mathbf{k}}/\partial k_{y}$ in an electronic system
does not contribute to the overall transverse motion (the observed Hall
current is purely from the anomalous velocity), while the initial atom
velocities $\partial \varepsilon _{\mathbf{k}}/\partial k_{y}$ dominate the
expansion process and the asymmetry of the column density should be small in
an atomic system. Therefore we need develop other techniques for the
observation of the intrinsic AHE.

Recently, the response of atomic density to a rotation of the trap has been
proposed for measuring quantum Hall conductivity for cold atoms based on the
well-known Str\u{e}da formula \cite{Streda} $\sigma =\partial n/\partial B$.
Here the rotation for atoms is equivalent to the magnetic field for
electrons. However, the Str\u{e}da formula does not apply to the AHE because
it is not quantized. Nevertheless, the density response to the rotation
still contains rich information about the intrinsic AHE. Consider a rotation
of the harmonic trap along the $z$ axis \cite{Haljan}, the Hamiltonian can
be written as
\begin{eqnarray}
H &=&\hbar ^{2}q^{2}/2m+\alpha \left( q_{x}\sigma _{y}-q_{y}\sigma
_{x}\right)  \notag \\
&&+h_{0}\sigma _{z}+m\left( \omega _{t}^{2}-\varpi ^{2}\right) r^{2}/2-\mu
\label{Hamrotation}
\end{eqnarray}%
in the rotation frame, where $\hbar \mathbf{q}=\hbar \mathbf{k}-m\varpi
\mathbf{\hat{z}}\times \mathbf{r}$ is the mechanical momentum of atoms \cite%
{Niu4}, $\varpi $ is the rotation frequency of the trap. The density of
atoms is
\begin{equation}
n\left( \mathbf{r}\right) =\left( 2\pi \right) ^{-2}\int d^{2}\mathbf{q}%
\left[ 1+m\varpi \Gamma _{z}/\hbar \right] f\left( \mathbf{q},\mathbf{r,}%
\varpi \right) ,  \label{density}
\end{equation}%
where $m\varpi \Gamma _{z}/\hbar $ is a correction to the well-known
constant density of states $1/\left( 2\pi \right) ^{2}$ in the presence of
nonzero Berry curvature fields and the rotation \cite{Xiao}. $f\left(
\mathbf{q},\mathbf{r,}\varpi \right) =1/\left( \exp \left[ \left(
\varepsilon _{\mathbf{m}}-\mu \left( \mathbf{r}\right) \right) /k_{B}T\right]
+1\right) $ is the Fermi Dirac distribution of atoms, where the energy of
atoms $\varepsilon _{\mathbf{m}}=\varepsilon _{\mathbf{q}}-M_{z}\varpi $
contains a correction $-M_{z}\varpi =-m\varpi \alpha ^{2}h_{0}/4\hbar \left(
\alpha ^{2}q^{2}+h_{0}^{2}\right) $, known as the magnetization energy for
electrons in the solid state \cite{Niu4}.

\begin{figure}[t]
\includegraphics[width=1.0\linewidth]{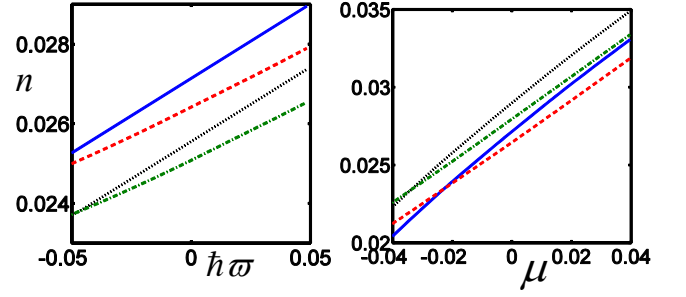} \vspace{-5pt} \vspace{%
-10pt}
\caption{(Color online) Plot of the atom density $n\left( \protect\varpi ,%
\protect\mu ,T\right) $ with respect to $\hbar \protect\varpi $ and $\protect%
\mu $. The units are the same as that in Fig. \protect\ref{TOF}. $%
h_{0}=0.05E_{R}$. $\protect\alpha ^{2}/\protect\gamma =1/3E_{R}$. Solid and
dotted: $T=2nK$; Dashed and dash dotted: $T=200nK$. (a) Solid and dashed: $%
\protect\mu =0$; Dotted and dash dotted: $\protect\mu =-0.01E_{R}$. (b)
Solid and dashed: $\protect\varpi =0$; Dotted and dash dotted: $\protect%
\varpi =0.05\protect\omega _{R}$.}
\label{densityplot}
\end{figure}

We assume the rotation frequency $\varpi $ of the trap is slowly increased
to keep the same temperature of the system. The local chemical potential can
be fixed by increasing $\omega _{t}$ with $\varpi $. The response of the
atom density to $\varpi $ is
\begin{equation}
\frac{\partial n}{\partial \varpi }=\int \frac{d^{2}\mathbf{q}}{\left( 2\pi
\right) ^{2}}\left[ \frac{m}{\hbar }\Gamma _{z}f+\left( 1+\frac{m\varpi
\Gamma _{z}}{\hbar }\right) \frac{\partial f}{\partial \mu }M_{z}\right]
\label{denderi}
\end{equation}%
for fixed $\mu $ and $T$. At $T=0$ and $\varpi =0$, Eq. (\ref{denderi})
reduces to%
\begin{equation}
\frac{\partial n}{\partial \varpi }=\frac{m}{\hbar }\int^{\mu }\frac{d^{2}%
\mathbf{q}}{\left( 2\pi \right) ^{2}}\Gamma _{z}\mathbf{+}\frac{M_{z}}{4\pi }%
\frac{1}{d\varepsilon _{\mathbf{q}}/dq^{2}|_{\varepsilon _{\mathbf{q}}=\mu }}%
.  \label{denderi2}
\end{equation}%
Here the first term is the anomalous Hall conductivity $\sigma _{xy}$ for
cold atoms. In the parameter region $\left\vert \mu \left( r\right)
\right\vert <h_{0}$ (\textit{i.e.}, the chemical potential lies in the band
gap), it yields $\sigma _{xy}=m\left[ 1-h_{0}/\sqrt{\alpha
^{2}q_{F}^{2}\left( r\right) +h_{0}^{2}}\right] /4\pi \hbar $, where the
Fermi wavevector $q_{F}\left( r\right) $ is obtained from $\varepsilon _{%
\mathbf{q}_{F}}=\mu \left( r\right) $. In the parameter region $\alpha
^{2}/\gamma \gg h_{0}$, $q_{F}^{2}\approx \alpha ^{2}/\gamma ^{2}$ and $%
\sigma _{xy}\approx m/4\pi \hbar $. The second term $\bar{\sigma}_{xy}$,
originating from the non-zero orbit magnetic moment $M_{z}$, is an
additional contribution to $\partial n/\partial \varpi $ that was missing in
the previous literature \cite{Xiao} for electron systems. Eq. (\ref{denderi2}%
) is a generalization of Str\u{e}da formula for the anomalous Hall effects.
By varying parameters and measuring the density response, we can extract
information not only about the anomalous Hall conductivity, but also the
magnetic moment that is generally hard to measure in solid state systems. In
the parameter region $\alpha ^{2}/\gamma \gg h_{0}$, $\bar{\sigma}%
_{xy}\approx \gamma mh_{0}/8\pi \hbar \alpha ^{2}\ll m/4\pi \hbar \approx
\sigma _{xy}$. Therefore $\sigma _{xy}$ dominates in Eq. (\ref{denderi}) in
this region, and the density response $\partial n/\partial \varpi $ yields a
rough measurement for the anomalous Hall conductivity. \newline

\begin{figure}[t]
\includegraphics[width=1.0\linewidth]{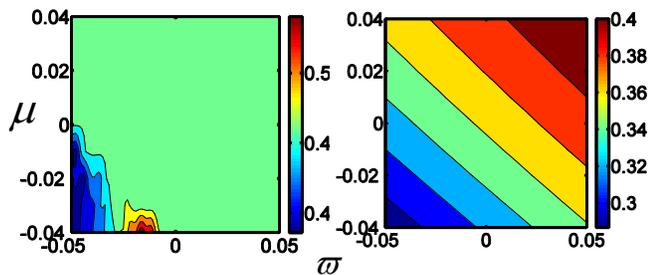} \vspace{-5pt} \vspace{-10pt}
\caption{(Color online) Plot of $\partial n\left( \protect\varpi ,\protect%
\mu ,T\right) /\partial \protect\varpi $ with respect to $\protect\varpi $
and $\protect\mu $. The unit of $\partial n\left( \protect\varpi ,\protect%
\mu ,T\right) /\partial \protect\varpi $ is taken to be $m/2\protect\pi %
\hbar $. The other units are the same as that in Fig. \protect\ref{TOF}. $%
h_{0}=0.05E_{R}$ and $\protect\alpha ^{2}/\protect\gamma =E_{R}/3$. (a) $T=2$
nK (b) $T=200$ nK.}
\label{fig:nder}
\end{figure}

We numerically calculate the density $n$ and density response $\partial
n/\partial \varpi $ as functions of the parameters $\left( \mu ,\varpi
,T\right) $, and plot them in Figs. \ref{densityplot} and \ref{fig:nder} for
$T=200$ nK and 2 nK. The presence of the harmonic trap changes the chemical
potential at $r$ by $-m\left( \omega _{t}^{2}-\varpi ^{2}\right) r^{2}/2$.
In a realistic experiment, the effective trapping frequency $\sqrt{\omega
_{t}^{2}-\varpi ^{2}}$ in the presence of rotation may be slightly different
from the initial trapping frequency $\omega _{t0}$ without rotation to keep
the same temperature of the system \cite{Ho}. This can be overcome by
comparing the densities at different spatial points $r$, $r^{\ast }$ such
that $\mu -\frac{1}{2}m\left( \omega _{t}^{2}-\varpi ^{2}\right) r^{\ast
2}=\mu -\frac{1}{2}m\omega _{t0}^{2}r^{2}$ to keep the same local chemical
potential \cite{Ho}. With this method, we can measure the density response
to the rotation with the fixed temperature and chemical potential. In
addition, the rotation of the system requires an asymmetric harmonic trap
\cite{Haljan}, which does not affect our results because it only change the
spatial positions for the measurement of the density change at a fixed
chemical potential through a different local chemical potential dependence $%
\mu \left( \mathbf{r}\right) =\mu -m[\left( \omega _{tx}^{2}-\varpi
^{2}\right) x^{2}+\left( \omega _{ty}^{2}-\varpi ^{2}\right) y^{2}]/2$. We
adopt a set of parameters: $\alpha ^{2}/\gamma =E_{R}/3=2\pi \hbar \times
2.8 $ KHz, $h_{0}=2\pi \hbar \times 425$ Hz, $\omega _{t0}=2\pi \times 50$
Hz, $\varpi _{\max }=0.05\omega _{R}=2\pi \times 425Hz$, $\omega _{t}\approx
\sqrt{\omega _{t0}^{2}+\varpi ^{2}}$, $\omega _{t}-\varpi =\omega
_{t0}^{2}/\left( \omega _{t}+\varpi \right) \approx 2\pi \times 3Hz$. At $%
r=0.1\lambda \omega _{R}/\pi \omega _{t0}$ $=4.2\mu m$, the chemical
potential $\mu $ changes by $0.01E_{R}$. From Figs. \ref{densityplot} and %
\ref{fig:nder}, we see a maximum density change at the order of $3\times
10^{7}$ cm$^{-2}$ can be observed with a rotation frequency $\varpi _{\max }$%
, which corresponds to about 10\% of the total density and can be observed
in a realistic experiment. The density variation at a medium temperature $%
T=200$ nK is at the same order as that at a low temperature $T=2$ nK. Note
that the multiple layer structure induced by the optical lattice confinement
along the $z$ direction can further enhance the signal.

In summary, we propose a scheme to create RSOC and PZF simultaneously for
cold atomic gases. We show that, by measuring the atomic density response to
a rotation of the trap, the intrinsic AHE can be observed for cold fermionic
atoms in a harmonic trap. We emphasize that the creation of RSOC and PZF
brings new opportunities for studying many topological phenomena, such as
chiral \textit{p}-wave superfluids, anomalous and spin Hall insulators,
\textit{etc}.

\textbf{Acknowledgments}: We thank Di Xiao and Qian Niu for helpful
discussion. This work is supported by the ARO (W911NF-09-1-0248) and
DARPA-YFA (N66001-10-1-4025).

\end{document}